\documentclass[nofootinbib,prd,]{revtex4}%
\usepackage{graphicx}
\usepackage{mathrsfs}
\usepackage{yfonts}
\usepackage{tipa}
\usepackage{bm}
\usepackage{bbm}
\usepackage{amsmath}
\usepackage{amssymb}
\usepackage{amsthm}
\usepackage{hyperref}
\usepackage{amsfonts}
\usepackage{mathtools}
\usepackage{latexsym}
\usepackage[greek,english]{babel}
\usepackage{booktabs}
\usepackage{xcolor}
\usepackage[iso-8859-7]{inputenc}%
\begin{document}
\title{Traversable Wormholes induced by Stress Energy Conservation: combining Casimir
Energy with a scalar field }
\author{Remo Garattini}
\email{remo.garattini@unibg.it}
\affiliation{Universit\`a degli Studi di Bergamo, Dipartimento di Ingegneria e Scienze
Applicate, Viale Marconi 5, 24044 Dalmine (Bergamo) Italy  and I.N.F.N.-
sezione di Milano, Milan, Italy.}
\author{Athanasios G. Tzikas}
\email{athanasios.tzikas@unibg.it}
\affiliation{Universit\`a degli Studi di Bergamo, Dipartimento di Ingegneria e Scienze
Applicate, Viale Marconi 5, 24044 Dalmine (Bergamo) Italy.}

\begin{abstract}
We investigate possible manifolds characterizing traversable wormholes in the presence of a scalar field minimally coupled to gravity, which has both kinetic and potential energy. The feature of traversability requires the violation of  the null energy condition, which, in turn, signals the existence of exotic matter with negative energy density. To achieve this, we introduce a hypothetical Casimir apparatus with plates positioned either at a parametrically fixed or radially varying distance. A consistent set of field equations requires the introduction of an auxiliary field composed solely of pressure terms, which we interpret as the gravitational back-reaction of the traversable wormhole to the original source. Interestingly, the only case that appears to avoid the need for such an auxiliary field involves a scalar field with potential energy, combined with a Casimir device with fixed plates.

\end{abstract}
\maketitle

\section{Introduction}

\label{sec:intro}

Wormholes are theoretical solutions of the Einstein Field Equations (EFE) that depict spacetime bridges connecting two distinct regions within the same universe or between different universes. They originate back in 1935 where Einstein and Rosen attempted to build a geometrical model of a particle playing the role of a bridge between two asymptotically flat spaces \cite{ER35}. Another significant 
 historical attempt was made by Wheeler in 1955 within the concept of spacetime foam \cite{Whe57}. According to his theory, quantum fluctuations of the metric would emerge near the Planck scale, resulting in a topology change that would lead in  tunneling between different regions of space (Planck-sized wormholes). Other notable attempts have been made over the years. For a comprehensive review, refer to \cite{Visser}. 
 
 Nevertheless, the aforementioned wormhole models are non-traversable by particles.  In 1973, independently, both Ellis \cite{Ellis} and  Bronnikov \cite{Bronnikov} demonstrated the possibility of the existence of a Traversable Wormhole (TW) by combining general relativity with a minimally coupled scalar field. The development of the scientific field continued with the publication of the Morris and Thorne paper in
1988 \cite{MT,MTY}. Adopting a more engineering-oriented
perspective for the wormhole, they first constructed the corresponding line
element of a TW and subsequently determined the matter required to construct such a spacetime. Their manifold can be described by a spherically symmetric metric that possesses two asymptotically flat regions. If we denote the proper radial distance by $l\,$, then the line element can be described by
\begin{equation} \label{in_metric}
\mathrm{d}s^{2}=-e^{2\Lambda(l)}\mathrm{d}t^{2}+\mathrm{d}l^{2}+r^{2} (l)  \left(  \mathrm{d}\theta^{2}+\sin^{2}\theta\ \mathrm{d}\phi
^{2}\right)  .
\end{equation}
The coordinate $l$ has a range defined in the interval $\left(  -\infty, +\infty\right)  \,$. The function $\Lambda(l)$ is an arbitrary function of $l$ and is known as the redshift function. To avoid event horizons (surfaces with
$e^{2\Lambda(l)}\rightarrow 0$), we require that $\Lambda(l)$ is finite
everywhere. The function $r (l)  $ must have the following
properties:%
\begin{equation}%
r ( l)  =\left\vert l\right\vert + \mathcal{O}\left(  1\right)    
(l\rightarrow\pm\infty ) \qquad \mathrm{and} \qquad
r_{0}=\min\left(  r\left(  l\right)  \right) \,. 
\end{equation}
The value $r_{0}$ represents the throat of the wormhole, corresponding to $l=0\,$. However, it is always possible to consider
Schwarzschild coordinates $\left(  t,r,\theta,\phi\right)  $ by substituting the
proper length through the following transformation%
\begin{equation}
l\left(  r\right)  =\pm\int_{r_{0}}^{r}\frac{\mathrm{d} \Tilde{r}}{\sqrt{1-b_{\pm
}(\Tilde{r})/\Tilde{r}}} \,.
\end{equation}
The function $b_{\pm}(r)$ is known as the shape function, where the  ``$\pm$''
refers to the upper universe ($+$) and the lower universe ($-$), respectively. To cover the entire spacetime, we now require two patches with the arbitrary functions  $b_{\pm}(r)$ and $\Lambda_{\pm}(r)\,$.  The line element \eqref{in_metric} then becomes
\begin{equation}
\mathrm{d}s^{2}=-e^{2\Lambda_{\pm}(r)}\mathrm{d}t^{2}+\frac{\mathrm{d}r^{2}%
}{1-b_{\pm}(r)/r}+r^{2}\left(  \mathrm{d}\theta^{2}+\sin^{2}\theta
\ \mathrm{d}\phi^{2}\right)  \,\label{metric00}%
\end{equation}
with $r\in [r_0,\infty)\,$. For simplicity, we can assume that $\Lambda_{+}(r)=\Lambda_{-}(r)$ and
$b_{+}(r)=b_{-}(r)\,$. The metric \eqref{metric00} then reduces to
\begin{equation} \label{metric}
\mathrm{d}s^{2}=-e^{2\Lambda(r)}\mathrm{d}t^{2}+\frac{\mathrm{d}r^{2}%
}{1-b(r)/r}+r^{2}\left(  \mathrm{d}\theta^{2}+\sin^{2}\theta\ \mathrm{d}%
\phi^{2}\right)  .
\end{equation}
In this coordinate system, the following condition%
\begin{equation}
1-b(r)/r>0\,
\end{equation}
must be satisfied, along with the conditions
\begin{equation} \label{flare}
b(r_{0})=r_{0}\qquad\mathrm{and}\qquad b^{\prime}(r_{0})\leq1 \,.
\end{equation}
The inequality in \eqref{flare} is called the flaring-out condition. The prime ($'$) denotes from now on differentiation with respect to the radial coordinate $r$.  In this work, we choose to employ Schwarzschild coordinates because it is convenient for solving equations. Let us also note that at $r=r_0\,$, the $g_{rr}$ component of the metric \eqref{metric}  diverges. This divergence though is apparent and clearly demonstrates a weakness of Schwarzschild coordinates in describing the behavior at the throat. For instance, the initial metric \eqref{in_metric} is everywhere regular, including the throat ($\ell=0$). Contrary to the proper distance  $\ell\,$, the radial coordinate $r$ does not directly represent any physical distance.

Unfortunately, the price one has to pay for the existence of a TW, is the requirement of an exotic matter as a constituent, which would violate the energy conditions. Although this may seem irrational at first, there are physical phenomena where such violations do occur. One well-known example is the Casimir experiment \cite{Cas48}, where two conducting plates placed very close to each other distort the zero-point energy, resulting this way in different pressures between the inner and outer walls. As a consequence, the plates attract each other since there are fewer virtual particles per unit volume between them than in the space outside the plates, indicating the presence of negative energy density in the gap. The Casimir effect provides a real experimental evidence that the energy conditions are occasionally violated by quantum effects. Under this framework, attempts have been made to utilize the Casimir source as exotic matter for a TW, leading to a solution known as Casimir wormhole \cite{CW}. But what happens when a Casimir wormhole is powered by a source of positive energy density? A partial attempt to answer this question
has been addressed in \cite{CCW}, where an electric field generated by a
point charge has been superposed to the original Casimir source. The Casimir source is described by  a Stress-Energy Tensor (SET) of the form
\begin{equation}
T_{\mu}^{\nu}|_{\mathrm{Casimir}} =\mathrm{diag}\left[  -\rho_{\mathrm{C}}\left(
r\right)  , \, p_{\mathrm{r,C}}\left(  r\right)  , \, p_{\mathrm{t,C}}\left(
r\right)  , \, p_{\mathrm{t,C}}\left(  r\right)  \right]   = -\frac{\hbar c \pi^2}{720 r^4} \mathrm{diag} \left[ -1,  3, -1, -1 \right]\, \label{casTmn}
\end{equation}
while an
electric field has a SET of the following form
\begin{equation}
T_{\mu}^{\nu}|_{\mathrm{EM}} =\mathrm{diag}\left[ - \rho_{\mathrm{E}}\left(
r\right)  , \, p_{\mathrm{r,E}}\left(  r\right)  , \, p_{\mathrm{t,E}}\left(
r\right)  , \, p_{\mathrm{t,E}}\left(  r\right)  \right]  =\frac{Q^{2}}{2\left(  4\pi\right)  ^{2}\varepsilon_{0}r^{4}
} \mathrm{diag} \left[ -1, -1, 1, 1 \right] \,.\label{TEM}
\end{equation}
The components of the electric energy density and the radial pressure  obey the following equation of state
\begin{equation}
\rho_{\mathrm{E}} \left(  r\right) =- p_{\mathrm{r},\mathrm{E}}\left(  r\right) .\label{prho}
\end{equation}
As a consequence, the Null Energy Condition (NEC) of the combined source is violated: 
\begin{equation}
\rho\left(  r\right)  +p_{\mathrm{r}}\left(  r\right)  =\rho_{\mathrm{C}}\left(  r\right)
+p_{\mathrm{r},\mathrm{C}}\left(  r\right)  +\rho_{\mathrm{E}}\left(  r\right)  +p_{\mathrm{r},\mathrm{E}}\left(  r\right)
=-\frac{4\hbar c\pi^{2}}{720r^{4}}<0 \,.
\end{equation}
To put it another way, the introduction of an electric field does not change the
violation of the NEC by means of the Casimir source. However, if one
attempts to extend this analysis to a scalar field replacing the electric field,
the answer may not be trivial. It is widely recognized that a scalar field of
opposite polarity can serve as a source for traversable wormholes even without
an additional Casimir source. Such an opposite polarity is commonly referred to as a
phantom field. The result is a well-known TW also dubbed as
Ellis-Bronniknov (EB) wormhole. 
 Although phantom fields represent a
valid source of exotic matter with negative energy density, one could
consider the effect of an ordinary scalar field combined with a Casimir
source. A description of this kind has been examined by S. Kim \cite{Kim}, who has described the combination of a source of exotic matter with a scalar field, whether it has a mass or not. Kim also discussed the possibility of having a
contribution from the effect of the back-reaction that could prevent the formation of a TW \cite{Kim1}. A particular relevant result has been obtained by C. Barcel\'{o} and M. Visser \cite{BV, BV1}, who examined the effect of a pure scalar field conformally coupled with a gravitational field. In these papers, the authors showed that an effective coupling constant is the key ingredient
to obtain a TW with positive scalar field sources. On the other hand, L.
Butcher \cite{Butcher} proved that every form of pure scalar field cannot be taken as a source for a TW.

In this paper, we are going to examine a
specific form of negative energy density coupled with an ordinary scalar field, considering cases with or without potential. The form of the negative energy density will be
described by the Casimir energy. We will find that, at this level, the
scalar field will not prevent the formation of a TW.   It is important to note that such an analysis is
not in contradiction with  Butcher's results because the boundary
conditions are different. To further proceed, we
will consider the following SET configuration
\begin{equation}
T_{\mu}^{\nu}=\mathrm{diag}\left[  -\rho\left(  r\right)  , \, p_{\mathrm{r}}\left(
r\right)  , \, p_{\mathrm{t}}\left(  r\right)  , \, p_{\mathrm{t}}\left(  r\right)
\right]\label{Tmn}
\end{equation}
where the quantities $\rho\left(  r\right)
$, $p_{\mathrm{r}}\left(  r\right)  $ and $p_{\mathrm{t}%
}(r)$ represent the total energy density, the total radial pressure
and the total tangential pressure of the anisotropic SET respectively. The energy-momentum conservation
is guaranteed by imposing the validity of the following differential equation%
\begin{equation}
p_{\mathrm{r}}^{\prime}\left(  r\right)  =\frac{2}{r}\left(  p_{\mathrm{t}%
}\left(  r\right)  -p_{\mathrm{r}}\left(  r\right)  \right)  -\left(
\rho\left(  r\right)  +p_{\mathrm{r}}\left(  r\right)  \right)  \Lambda
^{\prime}(r)\,. \label{set_cons}
\end{equation}
 We will consider a SET composed
by three parts: one part $T_{\mu\nu}^{\mathrm{Scalar}}$ resulting from the scalar field, another part $T_{\mu\nu}^{\mathrm{Casimir}}$ resulting from the Casimir
source  and the last part will be composed by a SET, whose form is
\begin{equation}
T_{\mu}^{\nu}|_{\mathrm{TW}}=\mathrm{diag}\left[ - \rho^{\mathrm{TW}} (r)
, \, p_{\mathrm{r}}^{\mathrm{TW}} (r)  , \, p_{\mathrm{t}}^{\mathrm{TW}} (r)
, \, p_{\mathrm{t}}^{\mathrm{TW}} (r)  \right]  .\label{TmnTW}
\end{equation}
This third piece is going to be the TW SET necessary to have
consistency and conservation of the total SET. Moreover, the term $T_{\mu\nu}^{\mathrm{TW}}$ will not be considered
as a contribution of an unknown matter field, rather it will be considered as
a kind of back-reaction of the TW to the matter source \cite{Kim1}.
Computationally, it is
convenient to introduce $T_{\mu\nu}^{\mathrm{TW}}$after having
computed the EFE with the Casimir and the scalar source. This is justified by
the fact that not every SET configuration needs the presence of $T_{\mu\nu
}^{\mathrm{TW}}$. The rest of the paper
is organized as follows: In section \ref{p2},  we analyze the structure of the SET for a massless scalar field without potential. Then we derive the solutions by enforcing SET conservation for such a field, considering two cases for the distance of the Casimir plates; one with fixed distance and another with radially variable distance. 
In section \ref{sec:massive}, we repeat the
same procedure as in the previous section, but this time for a massive scalar field with a
potential. We summarize and conclude in section
\ref{sec:concl}. Units in which $\hbar=c=k=1$ are used throughout the paper
and will be reintroduced whenever it is necessary.

\section{Casimir wormholes for massless scalar field}

\label{p2}The starting point of our analysis is the gravitational action governing our 
system  with the form of 
\begin{equation}
S=\int\mathrm{d}^{4}x\sqrt{-g}\left[  \frac{R}{2\kappa}-\frac{1}{2}(\nabla
\psi)^{2}-V(\psi)+\mathcal{L}_{\mathrm{C}}\right]  \, \label{action}
\end{equation} where $R$ is the Ricci scalar, $g$ is the determinant of the metric tensor,  $\kappa=8\pi G$,  $\psi$ is the scalar field, $\mathcal{L}
_{\mathrm{C}}$ is the Lagrangian of the Casimir source and $V(\psi)$ is the potential of the
scalar field. The potential is actually a functional of the radial coordinate, i.e.,
\begin{equation}
V(\psi)=V\left[\psi\left( r\right)\right]=V (r)\label{V}
\end{equation}
once we impose only radial dependence for the scalar field ($\psi=\psi(r)$).  We vary the action with respect
to the scalar field and the metric to derive the  differential
equations of motion. The variation with respect to the scalar field gives the following equation 
\begin{equation}
   \nabla^{2}
\psi=\frac{\mathrm{d}V(\psi)}{\mathrm{d} \psi} \,. \label{sc_field_eq}
\end{equation}
In this section, we consider the case where $V(\psi)=0\,$, which we refer to as the massless case. Then, the scalar field equation simplifies to $ \nabla^{2}
\psi = 0$ and can be cast
into the following component form
\begin{equation}
e^{2\Lambda(r)}r^{4}\left(  1-\frac{b(r)}{r}\right)  \psi^{\prime}
(r)^{2}=C\ (\mathrm{const.})\quad \mathrm{with} \quad  C > 0 \,.\label{eom_sc}
\end{equation}
Note that the case with a negative value for the integration constant $C$ ($C<0$) corresponds to the introduction of a phantom field.
The variation with respect to the metric yields the EFE, which are
\begin{align}
&  \frac{b^{\prime}(r)}{r^{2}}=\kappa \rho \left( r \right) \label{eom1}\\
&  \frac{2\Lambda^{\prime}(r)}{r}\left(  1-\frac{b(r)}{r}\right)  -\frac
{b(r)}{r^{3}}=\kappa p_{\mathrm{r}} \left( r \right) \label{pr(r)}\\
&  \left(  1-\frac{b(r)}{r}\right)  \left(  \Lambda^{\prime\prime}
(r)+\Lambda^{\prime}(r)^{2}+\frac{\Lambda^{\prime}(r)}{r}\right)  +\left(
\frac{b(r)-rb^{\prime}(r)}{2r^{3}}\right)  \left(  r\Lambda^{\prime
}(r)+1\right)  =\kappa p_{\mathrm{t}} \left( r \right)\, \label{pt(r)}
\end{align}
for the $tt-$, $rr-$ and $\theta \theta - $component respectively. The form of the Casimir components is given
by \eqref{casTmn}.
Before proceeding, it is essential  to emphasize that the forthcoming solutions imply also the conservation of the SET as represented by \eqref{set_cons}.
Regarding the Casimir apparatus, we  distinguish two different cases:
\begin{description}
\item[a)] the fixed Casimir plates case and 

\item[b)] the variable Casimir plates case.
\end{description}

\subsection{Fixed Casimir plates}

For the case \textbf{a)}, the Casimir walls are fixed at some distance $d$ and so the components of the total SET can be written as%
\begin{align}
\rho\left(  r\right)   &  =\frac{1}{2}\left(  1-\frac{b(r)}{r}\right)  \psi^{\prime}%
(r)^{2} -\frac{\hbar
c\pi^{2}}{720d^{4}}=\frac{Ce^{-2\Lambda(r)}}{2r^{4}}-\frac{r_{1}^{2}}{\kappa
d^{4}}\label{rhod}\\
p_{\mathrm{r}}\left( r\right)  &  = \frac{1}{2}\left(  1-\frac{b(r)}{r}\right)  \psi^{\prime}
(r)^{2} -\frac{3\hbar c\pi^{2}}{720d^{4}}=\frac{Ce^{-2\Lambda(r)}}{2r^{4}}-\frac{3r_{1}^{2}}{\kappa d^{4}} \label{prd}\\
p_{\mathrm{t}}\left( r\right) &=- \frac{1}{2}\left(  1-\frac{b(r)}{r}\right)  \psi^{\prime}%
(r)^{2} +\frac{\hbar c\pi^{2}
}{720d^{4}}=-\frac{Ce^{-2\Lambda(r)}}{2r^{4}}+\frac{r_{1}^{2}}{\kappa d^{4}}
\label{ptd}%
\end{align}
upon using \eqref{eom_sc} and after introducing the length scale
\begin{equation}
r_{1}^{2}=\frac{\hbar G}{c^{3}}\frac{\pi^{3}}{90}=\frac{\pi^{3}\ell
_{\mathrm{P}}^{2}}{90} \,.
\end{equation}
Inserting Eqs. $\left(  \ref{rhod},\ref{prd},\ref{ptd}\right)  $ into \eqref{set_cons}, one finds
\begin{equation}
-\frac{4\left(   r \Lambda\!^{\prime}\left(  r  \right)
+2\right)  r_{1}^{2}}{\kappa\,d^{4}r}=0
\end{equation}
that leads to
\begin{equation}
\Lambda(r)=-2\ln\left(  r\right)  +C_{1}\,.
\end{equation}
The integration constant $C_{1}$ is to be determined. We have two possibilities:

\begin{enumerate}
\item   $\Lambda(r_{0})=0$ ($C_{1}=2\ln  \left(r_0\right)   $), so that
\begin{equation}
\Lambda(r)=2\ln\left(  \frac{r_0}{r}\right)  
\end{equation}
and
\item  $\Lambda(\bar{r})=0$ ($C_{1}=2\ln\left(  \bar{r}\right)  $), meaning
\begin{equation}
\Lambda(r)=2\ln\left(  \frac{\bar{r}}{r}\right)  \label{Lambda}%
\end{equation}
where $\bar{r}$  is an outer cut-off boundary ($\bar{r}>r_{0}$) to be determined.
\end{enumerate}
The $\Lambda(r_{0})=0$ case gives no consistent solutions and so it will
be discarded. 
Plugging the
redshift function \eqref{Lambda} into \eqref{rhod}, one finds%
\begin{equation}
\rho\left(  r\right)  =\frac{C}{2\bar{r}^{4}}-\frac{r_{1}^{2}}{\kappa d^{4}%
}\gtreqless0\qquad \mathrm{if} \qquad C\gtreqless2\bar{r}^{4}r_{1}^{2}/\left(  \kappa
d^{4}\right)  .
\end{equation}
Despite  the simplicity of the energy density structure, it is better to
consider each case separately. We begin with the positive energy density case.

\subsubsection{Positive Energy Density}
\label{rhop} 

When $C>2\bar{r}^{4}r_{1}^{2}/\left(
\kappa d^{4}\right)  $, then
\begin{equation}
\rho\left(  r\right)  =\rho_{+}=\frac{C}{2\bar{r}^{4}}-\frac{r_{1}^{2}}{\kappa
d^{4}}>0
\end{equation}
and the first EFE \eqref{eom1} leads to%
\begin{equation}
b(r)=r_{0}+\frac{\kappa\rho_{+}}{3}\left(  r^{3}-r_{0}^{3}\right)
.\label{b(r)a}%
\end{equation}
Now $b(r)$ and $\Lambda(r)$ are exact solutions, but the redshift function has
the unpleasant feature that vanishes for large distances in a logarithmic way.  Therefore, we need a strategy to confine the whole solution in a well-defined range. To
further proceed, we insert \eqref{b(r)a} and \eqref{Lambda} into the second EFE \eqref{pr(r)},
retrieving
\begin{equation}
-\frac{4}{r^{2}}+\frac{3r_{0}}{r^{3}}-\frac{r_{1}^{2}}{d^{4}}+\frac{r_{0}%
^{3}r_{1}^{2}}{d^{4}r^{3}}+\frac{\kappa C}{2\bar{r}^{4}}-\frac{\kappa
Cr_{0}^{3}}{2r^{3}\bar{r}^{4}}=\frac{\kappa C}{2\bar{r}^{4}}-\frac{3r_{1}^{2}%
}{d^{4}}\,.\label{pr(r)a}%
\end{equation}
We have consistency if and only if the following equality%
\begin{equation}
C=\frac{2\bar{r}^{4}\left(  3r_{0}d^{4}-4\bar{r}d^{4}+r_{0}^{3}r_{1}%
^{2}+2r_{1}^{2}\bar{r}^{3}\right)  }{\kappa d^{4}r_{0}^{3}}\,\label{C}%
\end{equation}
is satisfied on the boundary $r=\bar{r}$. Plugging the value of $C$ into \eqref{pr(r)a}, one finds%
\begin{equation}
\left(  1-\frac{\bar{r}}{r}\right)  \left(  -\frac{4}{r^{2}}+\left(
1+\frac{\bar{r}}{r}+\frac{\bar{r}^{2}}{r^{2}}\right)  \frac{2r_{1}^{2}}{d^{4}%
}\right)  =0\,.
\end{equation}
Finally, the third EFE \eqref{pt(r)} reduces to%
\begin{equation}
\frac{4}{r^{2}}-\frac{9r_{0}}{2r^{3}}+\frac{r_{1}^{2}}{d^{4}}-\frac{3r_{0}%
^{3}r_{1}^{2}}{2d^{4}r^{3}}-\frac{\kappa C}{2\bar{r}^{4}}+\frac{3\kappa
Cr_{0}^{3}}{4r^{3}\bar{r}^{4}}=-\frac{\kappa C}{2\bar{r}^{4}}+\frac{r_{1}^{2}%
}{d^{4}}\,\label{pt(r)a}%
\end{equation}
and, with the help of \eqref{C}, we find%
\begin{equation}
\frac{4}{r^{2}}-\frac{6\bar{r}}{r^{3}}+\frac{3\bar{r}^{3}r_{1}^{2}}{d^{4}%
r^{3}}=0\,.
\end{equation}
The last equation gives the location of the boundary $\bar{r}$ in terms of
the plates separation and the Planck length%
\begin{equation}
\bar{r}=\sqrt{\frac{2}{3}}\frac{d^{2}}{r_{1}}\,.\label{BSize}%
\end{equation}
We can conclude that the EFE are satisfied and the consistency with the SET
conservation is preserved, at least on the boundary $r=\bar{r}$. Rearranging
\eqref{pr(r)a} and \eqref{pt(r)a} in terms of $\bar{r}$, one gets%
\begin{equation}
-\frac{4}{r^{2}}+\frac{8\bar{r}}{3r^{3}}+\frac{4}{3\bar{r}^{2}}=0\,
\end{equation}
for \eqref{pr(r)a} and%
\begin{equation}
\frac{4}{r^{2}}-\frac{4\bar{r}}{r^{3}}=0\,
\end{equation}
for \eqref{pt(r)a}. To have full consistency, we need to include the  tensor 
\eqref{TmnTW} which, in this case, is composed by the
following terms%
\begin{align}
p^{\mathrm{TW}}_{\mathrm{r}}\left(  r\right)   &  =\frac{4}{\kappa r^{2}}-\frac{8\bar{r}%
}{3\kappa r^{3}}-\frac{4}{3\kappa\bar{r}^{2}}\,\label{pr1(r)}\\
p^{\mathrm{TW}}_{\mathrm{t}}\left(  r\right)   &  =-\frac{4}{\kappa r^{2}}+\frac{4\bar{r}%
}{\kappa r^{3}}\,.\label{pt1(r)}%
\end{align}
Note that in $T_{\mu\nu}^{\mathrm{TW}}$ the energy density does not appear ($\rho^{\mathrm{TW}}(r)=0$).
It is important to observe that the value of the boundary $\bar{r}$ is also
useful to rearrange \eqref{C} in the following way%
\begin{equation}
C=\frac{2\bar{r}^{2}\left(  2r_{0}^{3}+9r_{0}\bar{r}^{2}-8\bar{r}^{3}\right)
}{3\kappa r_{0}^{3}}\,.\label{C1}%
\end{equation}
Since $C>0$, we have a constraint at the range of $\bar{r}$. One finds%
\begin{equation}
C>0\qquad\mathrm{if}\qquad\bar{r}<1.278r_{0}\,.
\end{equation}
With the help of \eqref{C1} and \eqref{BSize}, $b(r)$ can be cast into the
form%
\begin{equation}
b(r)=\frac{r^{3}}{r_{0}^{2}}-\frac{8\bar{r}r^{3}}{9r_{0}^{3}}+\frac{8\bar{r}%
}{9}\,.\label{b(r)ap}%
\end{equation}
Substituting next \eqref{C1} and \eqref{BSize} into $\rho_{+}$, we have to impose%
\begin{equation}
\bar{r}<\frac{9r_{0}}{8}\,.\label{BSize1}%
\end{equation}
Note also that it is possible to estimate the throat size thanks to
\eqref{BSize} and \eqref{BSize1}. We obtain%
\begin{equation}
r_{0}>\sqrt{\frac{2}{3}}\frac{8d^{2}}{9r_{1}}.
\end{equation}
It is easy to see that with the shape function described by \eqref{b(r)ap},
the flaring-out condition reads%
\begin{equation}
b^{\prime}(r_{0})=3\left(  1-\frac{8\bar{r}}{9r_{0}}\right)  <1\qquad
\Longrightarrow\qquad1<\frac{4\bar{r}}{3r_{0}}
\end{equation}
and is always satisfied.

\subsubsection{Vanishing Energy Density}

\label{rho0}In this part of the paper, we consider the case in which
$\rho\left(  r\right)  =0$. This can be realized when $C=2\bar{r}^{4}r_{1}%
^{2}/\left(  \kappa d^{4}\right)  $ and then the first EFE \eqref{eom1} leads to%
\begin{equation}
b(r)=r_{0}\,.\label{b(r)b}%
\end{equation}
The other SET components can be easily deduced. Indeed, we find%
\begin{equation}
\rho\left(  r\right)  =p_{\mathrm{t}}\left(  r\right)  =0
\end{equation}
and%
\begin{equation}
p_{\mathrm{r}}\left(  r\right)  =-\frac{2r_{1}^{2}}{\kappa d^{4}}\,.
\end{equation}
It is also easy to check that the SET is conserved, if the redshift function
is represented by \eqref{Lambda}. Plugging \eqref{Lambda} and $b(r)=r_{0}$
into the EFE \eqref{pr(r)} and \eqref{pt(r)}, one finds
\begin{equation}
-\frac{4}{r^{2}}+\frac{3r_{0}}{r^{3}}=-\frac{2r_{1}^{2}}{d^{4}}%
\,\label{pr(r)0}%
\end{equation}
for the EFE \eqref{pr(r)} and
\begin{equation}
\frac{4}{r^{2}}-\frac{9r_{0}}{2r^{3}}=0\,\label{pt(r)0}%
\end{equation}
for the EFE \eqref{pt(r)}. On the boundary $\bar{r}$, Eq.\eqref{pt(r)0} gives
\begin{equation}
\bar{r}=\frac{9r_{0}}{8}\,\label{rbrho0}%
\end{equation}
while from \eqref{pr(r)0} we find
\begin{equation}
r_{0}=\frac{8\sqrt{6}d^{2}}{27r_{1}}\,.
\end{equation}
Unfortunately, in this and the previous case, the throat is proportional to the square of the distance between the walls, implying the  existence of galactic-size wormholes, even if the distance $d$ is on the order of picometers. Furthermore, the EFE are satisfied on the boundary $\bar{r}$. Even in this
case, we need to introduce for consistency the SET $T_{\mu\nu}^{\mathrm{TW}}$  with the following components
\begin{align}
\rho^{\mathrm{TW}}\left(  r\right)   &  =0 \\
p^{\mathrm{TW}}_{\mathrm{r}}\left(  r\right)   &  =\frac{4}{\kappa r^{2}}-\frac{3r_{0}%
}{\kappa r^{3}}-\frac{2r_{1}^{2}}{\kappa d^{4}}\, \\
p^{\mathrm{TW}}_{\mathrm{t}}\left(  r\right)   &  =-\frac{4}{\kappa r^{2}}+\frac{9r_{0}%
}{2\kappa r^{3}}\,.
\end{align}
Let us mention here that it is also possible to obtain a vanishing energy
density with the help of the following redshift function
\begin{equation}
\Lambda(r)=\frac{1}{2}\ln\left(  \frac{C\kappa d^{4}}{2r_{1}^{2}r^{4}}\right)
\,\label{L_rho0}%
\end{equation}
which has the same structure as that in  \eqref{Lambda}.

\subsubsection{Negative Energy Density}

\label{rhom}

The negative density $\rho\left(  r\right)  <0$ is present when $C<2\bar
{r}^{4}r_{1}^{2}/\left(  \kappa d^{4}\right)  $, so that%
\begin{equation}
\rho_{-}=\frac{r_{1}^{2}}{\kappa d^{4}}-\frac{C}{2\bar{r}^{4}}>0 \,.
\end{equation} 
Then, the first EFE \eqref{eom1} will lead to%
\begin{equation}
b(r)=r_{0}-\frac{\kappa\rho_{-}}{3}\left(  r^{3}-r_{0}^{3}\right)
.\label{b(r)c}%
\end{equation}
For this profile, there exists $r=\bar{r}$ such that $b(\bar{r})=0$ and this is
located at%
\begin{equation}
\bar{r}=r_{0}\sqrt[3]{1+\frac{3}{\kappa\rho_{-}r_{0}^{2}}}\,.
\end{equation}
To further proceed, we substitute \eqref{b(r)c} and \eqref{Lambda} into the second EFE
\eqref{pr(r)}, retrieving%
\begin{equation}
-\frac{4}{r^{2}}+\frac{3r_{0}}{r^{3}}-\frac{r_{1}^{2}}{d^{4}}+\frac{r_{0}%
^{3}r_{1}^{2}}{d^{4}r^{3}}+\frac{\kappa C}{2\bar{r}^{4}}-\frac{\kappa
Cr_{0}^{3}}{2r^{3}\bar{r}^{4}}=\frac{\kappa C}{2\bar{r}^{4}}-\frac{3r_{1}^{2}%
}{d^{4}}\,.\label{puz}%
\end{equation}
We have consistency if and only if the above equality is satisfied at the
throat. This requirement provides a similar expression for the constant $C$
with \eqref{C}. Inserting this value of $C$ into \eqref{puz}, one finds%
\begin{equation}
-\frac{4}{r^{2}}+\frac{4\bar{r}}{r^{3}}+\frac{2r_{1}^{2}}{d^{4}}-\frac
{2\bar{r}^{3}r_{1}^{2}}{d^{4}r^{3}}=0\,.
\end{equation}
Finally, the third EFE \eqref{pt(r)} reduces to%
\begin{equation}
\frac{4}{r^{2}}-\frac{9r_{0}}{2r^{3}}+\frac{r_{1}^{2}}{d^{4}}-\frac{3r_{0}%
^{3}r_{1}^{2}}{2d^{4}r^{3}}-\frac{\kappa C}{2\bar{r}^{4}}+\frac{3\kappa
Cr_{0}^{3}}{4r^{3}\bar{r}^{4}}=-\frac{\kappa C}{2\bar{r}^{4}}+\frac{r_{1}^{2}%
}{d^{4}}\,
\end{equation}
and, with the help of $C$, we find%
\begin{equation}
\frac{4}{r^{2}}-\frac{6\bar{r}}{r^{3}}+\frac{3\bar{r}^{3}r_{1}^{2}}{d^{4}%
r^{3}}=0\,.
\end{equation}
On the boundary $r=\bar{r}$, we obtain%
\begin{equation}
\bar{r}=\sqrt{\frac{2}{3}}\frac{d^{2}}{r_{1}}\,.\label{rbneg}%
\end{equation}
Plugging \eqref{C} and \eqref{rbneg} into \eqref{b(r)c}, we find%
\begin{equation}
b(r)=\left(  \frac{1}{r_{0}^{2}}-\frac{8\bar{r}}{9r_{0}^{3}}\right)
r^{3}+\frac{8\bar{r}}{9}\,.\label{b(r)cm}%
\end{equation}
The full consistency is obtained with the help of $\left(  \ref{TmnTW}\right)
$ where%
\begin{align}
\rho^{\mathrm{TW}}\left(  r\right)   &  =0\\
p^{\mathrm{TW}}_{\mathrm{r}}\left(  r\right)   &  =-\frac{4}{\kappa r^{2}}+\frac{8\bar{r}%
}{3\kappa r^{3}}+\frac{4}{3\kappa\bar{r}^{2}}\,\\
p^{\mathrm{TW}}_{\mathrm{t}}\left(  r\right)   &  =\frac{4}{\kappa r^{2}}-\frac{6\bar{r}%
}{\kappa r^{3}}+\frac{3\bar{r}^{3}r_{1}^{2}}{\kappa d^{4}r^{3}}\,.
\end{align}
In terms of $r_{0}$ and $\bar{r}$, we can rearrange the value of $\rho_{-}$ as
\begin{equation}
\rho_{-}=\frac{8\bar{r}-9r_{0}}{3\kappa r_{0}^{3}}\,.\label{rho-}%
\end{equation}
It is easy to see that from \eqref{b(r)cm}, the flaring-out condition becomes%
\begin{equation}
b^{\prime}(r_{0})<1\qquad\Longleftrightarrow\qquad1<\frac{4\bar{r}}{3r_{0}}%
\end{equation}
and is compatible with \eqref{rho-}. From $C>0$ and from \eqref{rho-}, we get
the validity range for the cut-off boundary
\begin{equation}
1.125r_{0}<\bar{r}<1.278r_{0}\,.
\end{equation}

\subsection{Variable Casimir plates}

For the case \textbf{b)}, we know that the total SET is formed by the following components
\begin{align}
\rho\left(  r\right)   &  =\frac{Ce^{-2\Lambda(r)}}{2r^{4}}-\frac{r_{1}^{2}%
}{\kappa r^{4}} \,\label{rhor}\\
p_{\mathrm{r}}\left(  r\right)   &  =\frac{Ce^{-2\Lambda(r)}}{2r^{4}}%
-\frac{3r_{1}^{2}}{\kappa r^{4}} \, \label{prr} \\ 
p_{\mathrm{t}}\left(  r\right) & =-\frac{Ce^{-2\Lambda(r)}}{2r^{4}}+\frac
{r_{1}^{2}}{\kappa r^{4}} \,. \label{ptr}
\end{align}
Inserting Eqs.$\left(  \ref{rhor},\ref{prr},\ref{ptr}\right)  $ into
\eqref{set_cons}, one finds%
\begin{equation}
\frac{4 \left(  1-\Lambda\!^{\prime}\left(  r\right)  \right)  r_{1}^{2}%
}{\kappa r^{4}}=0 \label{Lambdar}%
\end{equation}
leading to%
\begin{equation}
\Lambda(r)=\ln\left(  r\right)  + C_{2} \,
\end{equation}
where $C_{2}$ is an integration constant to be determined. Even in the
variable case, we have two possibilities:

\begin{enumerate}
\item $\Lambda(r_{0})=0$ ($C_2=-\ln\left(  r_0\right)$), that means%
\begin{equation}
\Lambda(r)=\ln\left(  \frac{r}{r_{0}}\right)
\end{equation}
and
\item $\Lambda(\bar{r})=0$ ($C_2=-\ln\left(  \bar{r}\right)$), namely%
\begin{equation}
\Lambda(\bar{r})=\ln\left(  \frac{r}{\bar{r}}\right) \,.  \label{Lrrb}%
\end{equation}
\end{enumerate}
Once again, the $\Lambda(r_{0})=0$ case does not provide any consistent solution and so it will
be discarded. On the contrary, inserting \eqref{Lrrb} into \eqref{rhor}, leads
to%
\begin{equation}
\rho\left(  r\right)  =\frac{C\bar{r}^{2}}{2r^{6}}-\frac{r_{1}^{2}}{\kappa
r^{4}}\gtreqless0\qquad\mathrm{if}\qquad\sqrt{\frac{C\kappa\bar{r}^{2}}%
{2r_{1}^{2}}}\frac{\bar{r}}{r_{1}}\gtreqless r\,.
\end{equation}
Substituting next the redshift function \eqref{Lrrb} in \eqref{eom1}, one
finds%
\begin{equation}
b\left(  r\right)  =r_{0}+\frac{\bar{r}^{2}C\kappa}{6}\left(  \frac{1}%
{r_{0}^{3}}-\frac{1}{r^{3}}\right)  +r_{1}^{2}\left(  \frac{1}{r}-\frac
{1}{r_{0}}\right)  \,.\label{b(r)var}%
\end{equation}
With the helpf of \eqref{b(r)var} and \eqref{Lrrb} and from the second EFE
\eqref{pr(r)}, we obtain
\begin{equation}
\frac{2}{r^{2}}-\frac{3r_{0}}{r^{3}}-\frac{\bar{r}^{2}C\kappa}{2r^{3}r_{0}%
^{3}}+\frac{3r_{1}^{2}}{r^{3}r_{0}}+\frac{\bar{r}^{2}C\kappa}{2r^{6}}%
-\frac{3r_{1}^{2}}{r^{4}}=\frac{\kappa C\bar{r}^{2}}{2r^{6}}-\frac{3r_{1}^{2}%
}{r^{4}}\,.\label{pr(r)var}%
\end{equation}
Evaluating the previous equation on the boundary $\bar{r}$, we find the
value of $C$%
\begin{equation}
C=\frac{2r_{0}^{2}}{\kappa\bar{r}^{2}}\left(  2\bar{r}r_{0}-3r_{0}^{2}%
+3r_{1}^{2}\right)  \,.\label{Cvar}%
\end{equation}
Since $C$ must be positive, we have an additional constraint%
\begin{equation}
\bar{r}>\frac{3\left(  r_{0}^{2}-r_{1}^{2}\right)  }{2r_{0}}>0\qquad
\Longrightarrow\qquad r_{0}>r_{1}\,.\label{constr1}%
\end{equation}
Plugging \eqref{Cvar} into \eqref{pr(r)var}, we find%
\begin{equation}
\frac{2\left(  r-\bar{r}\right)  }{r^{3}}=0\,
\end{equation}
which is satisfied on the boundary as it should be. The last equation we need
to examine is the third EFE \eqref{pt(r)}. We simply obtain
\begin{equation}
\frac{1}{r^{2}}=0 \,\label{Inc}%
\end{equation}
which is meaningless. However, we have to remember that we also need to consider  $T_{\mu\nu}^{\mathrm{TW}}$ to cure such an inconsistency which, in this
case, consists of the following components
\begin{align}
\rho^{\mathrm{TW}}\left(  r\right)   &  =0 \,\\
p^{\mathrm{TW}}_{\mathrm{r}}\left(  r\right)   &  =-\frac{2\left(  r-\bar{r}\right)
}{\kappa r^{3}}\,\\
p^{\mathrm{TW}}_{\mathrm{t}}\left(  r\right)   &  =-\frac{1}{\kappa r^{2}}\,.
\end{align}
We can rearrange the shape function with the help of \eqref{Cvar}. One then
gets%
\begin{equation}
b\left(  r\right)  =\frac{r_{0}^{4}}{r^{3}}-\frac{2\bar{r}\,r_{0}^{3}}{3r^{3}%
}+\frac{2\bar{r}}{3}-\frac{r_{0}^{2}r_{1}^{2}}{r^{3}}+\frac{r_{1}^{2}}%
{r}\,.\label{b(r)var1}%
\end{equation}
It remains to check the validity of the flaring-out condition ($b^{\prime
}\left(  r_{0}\right)  <1$). On the one hand, from \eqref{b(r)var1} we obtain%
\begin{equation}
b^{\prime}\left(  r\right)  =-3\frac{r_{0}^{4}}{r^{4}}+\frac{2\bar{r}%
\,r_{0}^{3}}{r^{4}}+3\frac{r_{0}^{2}r_{1}^{2}}{r^{4}}-\frac{r_{1}^{2}}{r^{2}}%
\end{equation}
and so%
\begin{equation}
b^{\prime}\left(  r_{0}\right)  =-3+\frac{2\bar{r}}{r_{0}}+\frac{2r_{1}^{2}%
}{r_{0}^{2}}<1\qquad\Longrightarrow\qquad\bar{r}<2r_{0}-\frac{r_{1}^{2}}%
{r_{0}}\,.\label{Flare}%
\end{equation}
On the other hand, the inequality $\left(  \ref{constr1}\right)  $ must also
be satisfied and hence
\begin{equation}
2r_{0}-\frac{r_{1}^{2}}{r_{0}}>\bar{r}>\frac{3r_{0}}{2}-\frac{3r_{1}^{2}%
}{2r_{0}}\,.
\end{equation}
Since $r_{0}>r_{1}>0$, we can assume that there exists
$\alpha>1$ such that $r_{0}=\alpha r_{1}$. Therefore, the last inequality
becomes%
\begin{equation}
\frac{r_{1}}{\alpha}\left(  2\alpha^{2}-1\right)  >\bar{r}>\frac{3r_{1}%
}{2\alpha}\left(  \alpha^{2}-1\right)  .
\end{equation}

\section{Casimir wormholes with a non-vanishing potential for a scalar field}
\label{sec:massive}

In this section we proceed with a non-vanishing potential for the scalar field
according to \eqref{V}. This case is referred to as the massive case. Now the field equation \eqref{sc_field_eq} becomes
\begin{equation}
\left(  1-\frac{b(r)}{r}\right)  \left[  \psi^{\prime\prime
}(r)+\left(  \frac{2}{r}+\Lambda^{\prime}(r)\right)  \psi^{\prime}(r)\right]
+\left(  \frac{b(r)-rb^{\prime}(r)}{2r^{2}}\right)  \psi^{\prime}%
(r)=\frac{V^{\prime}(r)}{\psi^{\prime}(r)} \,\label{scalar_eom}%
\end{equation}
where we have assumed that%
\begin{equation}
\frac{\mathrm{d}V(\psi)}{\mathrm{d}\psi}=\frac{V^{\prime}(r)}{\psi^{\prime
}(r)}\qquad \mathrm{with} \qquad \psi^{\prime}(r)\neq 0 \,.
\end{equation}
The EFE are given by \eqref{eom1}-\eqref{pt(r)} with the modification
\begin{align}
\rho\left( r\right) &  =\frac{1}{2}\left(  1-\frac{b(r)}{r}\right)  \psi^{\prime}(r)^{2}+V(r)+\rho_{\mathrm{C}} \left( r\right)\,\label{t_press2}\\
p_{\mathrm{r}}\left( r\right) &  =\frac{1}{2}\left(  1-\frac{b(r)}{r}\right)  \psi
^{\prime}(r)^{2}-V(r)+p_{\mathrm{r,C}} \left( r\right)\,\\
p_{\mathrm{t}}\left( r\right) &  =-\frac{1}{2}\left(  1-\frac{b(r)}{r}\right)
\psi^{\prime}(r)^{2}-V(r)+p_{\mathrm{t,C}}\left( r\right).
\end{align}
 As before, we examine two different cases; fixed and variable Casimir plates.

\subsection{Fixed Casimir plates}
\label{subsec:F}

In this case, the components of the SET are
\begin{align}
\rho\left( r\right) &  =\frac{1}{2}\left(  1-\frac{b(r)}{r}\right)  \psi^{\prime}%
(r)^{2}+V(r)-\frac{r_{1}^{2}}{\kappa d^{4}}\,\label{t_press3}\\
p_{\mathrm{r}}\left( r\right) &  =\frac{1}{2}\left(  1-\frac{b(r)}{r}\right)  \psi
^{\prime}(r)^{2}-V(r)-\frac{3r_{1}^{2}}{\kappa d^{4}}\,\\
p_{\mathrm{t}}\left( r\right) &  =-\frac{1}{2}\left(  1-\frac{b(r)}{r}\right)
\psi^{\prime}(r)^{2}-V(r)+\frac{r_{1}^{2}}{\kappa d^{4}}\,.
\end{align}
and its conservation \eqref{set_cons}  gives the following equation
\begin{equation}
\left(  1-\frac{b(r)}{r}\right)  \left[  \psi^{\prime\prime}(r)\psi^{\prime
}(r)+\left(  \frac{2}{r}+\Lambda^{\prime}(r)\right)  \psi^{\prime}%
(r)^{2}\right]  +\left(  \frac{b(r)-rb^{\prime}(r)}{2r^{2}}\right)
\psi^{\prime}(r)^{2}-V^{\prime}(r)=\frac{4r_{1}^{2}}{\kappa d^{4}}\left(
\frac{2}{r}+\Lambda^{\prime}(r)\right)  .\label{SET3A}%
\end{equation}
If $\psi(r)$ is a solution of the scalar field equation \eqref{scalar_eom},
then the left-hand side of \eqref{SET3A} is identically zero. As a result, the
SET conservation gives the solution for the redshift function
\begin{equation}
\label{L3}\Lambda(r) = C_{3} - 2\ln r \,
\end{equation}
where the integration constant $C_3$ is yet to be determined. Next, by adding \eqref{eom1}
and \eqref{pt(r)}, we get
\begin{equation}
\frac{b^{\prime}(r)}{r^{2}} + \left(  1 - \frac{b(r)}{r} \right)  \left(
\Lambda^{\prime\prime}(r) + \Lambda^{\prime}(r)^{2} + \frac{\Lambda^{\prime
}(r)}{r} \right)  + \left(  \frac{b(r)-r b^{\prime}(r)}{2r^{3}} \right)
\left(  r \Lambda^{\prime}(r) + 1 \right) =0
\end{equation}
and by inserting \eqref{L3}, we solve over the shape function which is
calculated to be
\begin{equation}
\label{b3}b(r) = \frac{4r}{3} - \frac{r^{3}}{3r_{0}^{2}} \,.
\end{equation}
The shape function does not vanish at infinity and so we need an external
cut-off boundary ($r=\bar{r}$) where $b(\bar r)=0$. Such a boundary is simply
given by $\bar r= 2r_{0}$, leaving us with a physically meaningful inner region
for the traversable wormhole, i.e., $r \in[r_{0},\, 2r_{0}]$. Let us note that, in this section and the previous one, the solutions have been derived within a physical range bounded by an external boundary $\bar{r}$. Consequently, spacetime is divided into two distinct regions; an inner region representing the traversable wormhole and an outer region representing asymptotically flat space. 
The task of merging the outer region with the inner region is a crucial step and requires  the use of thin-shell techniques\footnote{Thin-shell techniques refer to methods used to represent a boundary surface, or a thin shell, within a spacetime that separates two distinct regions \cite{Isr66,Sushkov}. These techniques can indeed be used to ``glue'' together different spacetimes by providing a mathematical framework for describing the interface between them.}. Such a task though is beyond the scope of this paper and represents a potential area for future research.

Continuing with the calculation of the flaring-out condition, we find that $b^{\prime}(r_{0})=1/3<1$, indicating that the condition is always satisfied.
Being that said, the integration constant $C_{3}$ is chosen to be $C_{3}%
=2\ln\bar r$ so that
\begin{equation}
\label{L3b}\Lambda(r) = 2\ln\left(  \frac{\bar r}{r} \right)  .
\end{equation}
One sees that the redshift function also vanishes at the boundary $\bar{r}$.
It also is interesting to observe that Eqs. \eqref{b3} and 
\eqref{L3b} look like an Absurdly Benign Traversable Wormhole (ABTW) \cite{GABTW} defined by 
\begin{equation}
b(r)=\left\{
\begin{array}
[c]{cc}%
r_0 \left( 1-\mu \left( r-r_{0}\right) \right) ^{2}  \qquad & r_{0}\leq r < \bar{r}  \\
& \\
0   &  r\geq \bar{r}\,
\end{array}
\right.
\end{equation}
with $\Lambda(r)=0$ everywhere and $\bar{r}=r_{0}+1/\mu  \,$. Indeed, by rearranging \eqref{b3}, one finds that, near the throat
\begin{equation}
b(r)\simeq 2r\left( 1-\frac{r}{2r_{0}}\right) .
\end{equation}%
Thus, we can define a Quasi-ABTW described by%
\begin{equation}
b(r)=\left\{
\begin{array}
[c]{ccc}%
2r\left( 1-\frac{r}{2r_{0}}\right) \,, \qquad &  0 \leq \Lambda(r) \leq \ln 4 \,, \qquad & r_{0}\leq r <  \bar{r}  \\
& \\
0  \,, & \Lambda(r)=0 \,,& r\geq \bar{r}\,
\end{array}
\right.
\end{equation}
with $\bar{r}=2r_0\,$.
In this case, the energy density does not vanish for $r \geq \bar{r}$ but instead becomes constant. 
We proceed further by subtracting \eqref{pr(r)} from \eqref{eom1}, getting
\begin{equation}
\frac{b^{\prime}(r)}{r^{2}} - \frac{2\Lambda^{\prime}(r)}{r} \left(  1-
\frac{b(r)}{r} \right)  + \frac{b(r)}{r^{3}} = 2\kappa V(r) + \frac{2r_{1}%
^{2}}{d^{4}} \,
\end{equation}
and with the help of \eqref{L3b} and \eqref{b3}, we solve over the potential
\begin{equation}
\label{V1}V(r) = \frac{2}{3\kappa r^{2}} - \frac{r_{1}^{2}}{\kappa d^{4}} \,.
\end{equation}
To check the violation of the Weak Enrgy Condition (WEC), we calculate the energy density
\eqref{t_press3} as
\begin{equation}
\rho\left( r\right) = \frac{b^{\prime}(r)}{\kappa r^{2}} = \frac{4}{3\kappa r^{2}} -
\frac{1}{\kappa r_{0}^{2}} \,.
\end{equation}
The negative value of the energy density is guarantied if $r > 2r_{0}/\sqrt
{3}$. Hence, there is a range $r \in[r_{0}, \, 2r_{0}/\sqrt{3}]$ where the WEC
is satisfied. On the contrary, the NEC is always violated because
\begin{equation}
\rho\left( r\right) + p_{r}\left( r\right) = \frac{4}{3\kappa r^{2}} - \frac{2}{\kappa r_{0}^{2}} <0
\qquad\mathrm{for} \qquad r \geq r_{0} \,.
\end{equation}
Finally, we may calculate the function of the scalar field by
inserting the forms we found for $b(r)$, $\Lambda(r)$ and $V(r)$ in any of the
EFE. The result is the following differential equation%
\begin{align}
&  \frac{r^{3}\kappa\left(  \left(  \psi^{\prime}(r)\right)  ^{2}\right)
^{\prime}\left(  r^{2}-r_{0}^{2}\right)  +2r^{4}\left(  \psi^{\prime}\!\left(
r\right)  \right)  ^{2}\kappa+8r_{0}^{2}}{6\psi^{\prime}\!\left(  r\right)
\kappa\,r^{3}r_{0}^{2}}\nonumber\\
&  =\frac{r^{3}\kappa y^{\prime}(r)\left(  r^{2}-r_{0}^{2}\right)
+2r^{4}y\!\left(  r\right)  \kappa+8r_{0}^{2}}{6\psi^{\prime}\!\left(
r\right)  \kappa\,r^{3}r_{0}^{2}}=0 \,
\end{align}
where $\left(  \psi^{\prime}\!\left(
r\right)  \right)  ^{2}=y\!\left(  r\right) $ and the solution for the scalar field is calculated to be%
\begin{align}
\psi(r) &  =\int_{r_{0}}^{r}\sqrt{\frac{C\kappa\,x^{2}+4r_{0}^{2}}%
{\kappa\,x^{2}\left(  x^{2}-r_{0}^{2}\right)  }}\mathrm{d}x=\frac{\pi}{2}%
+\arctan\!\left(  \frac{\sqrt{r^{4}+3r^{2}r_{0}^{2}-4r_{0}^{4}}}{2r_{0}^{2}%
}\right)  +\frac{\ln\!\left(  2\sqrt{r^{2}+4r_{0}^{2}}\sqrt{r^{2}-r_{0}^{2}%
}+2r^{2}+3r_{0}^{2}\right)  }{4}\\
&  -\arctan\!\left(  \frac{4r^{2}+6r_{0}^{2}}{3\sqrt{r^{4}+3r^{2}r_{0}%
^{2}-4r_{0}^{4}}}\right)  -\frac{\ln\!\left(  2r^{2}+3r_{0}^{2}-2\sqrt
{r^{2}+4r_{0}^{2}}\,\sqrt{r^{2}-r_{0}^{2}}\right)  }{4} \,.
\end{align}
Close to the throat, one finds%
\begin{equation}
\psi(r)\simeq\frac{\sqrt{10}}{\sqrt{r_{0}}}\sqrt{r-r_{0}}+ \mathcal{O}\! \left(
r-r_{0}\right)  .
\end{equation}
It is imperative to observe that this solution represents the only case in which
$T_{\mu\nu}^{\mathrm{TW}}$ does not come into play.

\subsection{Variable Casimir plates}

In this last case, the components of the SET are given by
\begin{align}
\label{t_press4}\rho \left( r\right)  &  = \frac{1}{2} \left(  1 - \frac{b(r)}{r} \right)
\psi^{\prime} \left( r\right)^{2} + V(r) - \frac{r_{1}^{2}}{\kappa r^{4}} \,\\
p_{\mathrm{r}} \left( r\right)  &  = \frac{1}{2} \left(  1 - \frac{b(r)}{r} \right)
\psi^{\prime}(r)^{2} - V(r) - \frac{3r_{1}^{2}}{\kappa r^{4}} \,\\
p_{\mathrm{t}} \left( r\right)  &  = - \frac{1}{2} \left(  1 - \frac{b(r)}{r} \right)
\psi^{\prime}(r)^{2} -V(r) + \frac{r_{1}^{2}}{\kappa r^{4}} 
\end{align}
and its conservation gives the following equation
\begin{equation}
\left(  1 - \frac{b(r)}{r} \right)  \left[  \psi^{\prime\prime}(r)
\psi^{\prime}(r) + \left(  \frac{2}{r}+ \Lambda^{\prime}(r) \right)
\psi^{\prime}(r)^{2} \right]  + \left(  \frac{ b(r) -r b^{\prime}(r) }{2r^{2}}
\right)  \psi^{\prime}(r)^{2} -V^{\prime}(r) = - \frac{12 r_{1}^{2}}{\kappa
r^{5}} + \frac{4r_{1}^{2}}{\kappa r^{4}}\left(  \frac{2}{r}+\Lambda^{\prime
}(r) \right)  .\label{SET3B}%
\end{equation}
Even in this case, if $\psi(r)$ is a solution of the scalar field equation
\eqref{scalar_eom}, then the left-hand side of \eqref{SET3B} is identically
zero. Therefore, the SET gives the following solution for the redshift function
\begin{equation}
\label{L4}\Lambda(r) = \ln r + C_{4} \,
\end{equation}
where the integration constant can be identified as $C_{4}=-\ln\bar{r}$ with
$r=\bar r$ being again a cut-off boundary for the wormhole. We have
to solve%
\begin{equation}
\left(  1-\frac{b(r)}{r}\right)  \left[  \frac{1}{2}\frac{\mathrm{d}}{\mathrm{d}r}\left(
\psi^{\prime}(r)\right)  ^{2}+\left(  \frac{2}{r}+\Lambda^{\prime}(r)\right)
\psi^{\prime}(r)^{2}\right]  +\left(  \frac{b(r)-rb^{\prime}(r)}{2r^{2}%
}\right)  \psi^{\prime}(r)^{2}-V^{\prime}(r)=0 \,
\end{equation}
which can be rearranged to give%
\begin{equation}
\frac{\mathrm{d}}{\mathrm{d}r}\left[  \frac{1}{2}\left(  1-\frac{b(r)}{r}\right)  \left(
\psi^{\prime}(r)\right)  ^{2}\right]  +\left[  \frac{3}{r}\left(
1-\frac{b(r)}{r}\right)  \psi^{\prime}(r)^{2}\right]  -V^{\prime}(r)=0 \,.
\end{equation}
This equation can be solved as a function of $V^{\prime}(r)$. Indeed, the
previous equation can be cast into the form%
\begin{gather}
y(r)= \frac{1}{2}\left(  1-\frac{b(r)}{r}\right)  \left(  \psi^{\prime}(r)\right)  ^{2},\\
y^{\prime}(r)+\frac{6}{r}y(r)-V^{\prime}(r)=0 \,
\end{gather}
with
\begin{equation}
y(r)=\left(  1-\frac{b(r)}{r}\right)  \left(  \psi^{\prime}(r)\right)
^{2}=\frac{1}{r^{6}}\left(  C_5+2\int \mathrm{d}r \ r^6 V^{\prime}(r)\right)  .
\end{equation}
Isolating $\psi^{\prime}(r)$, we obtain%
\begin{equation}
\psi(r)=\int \mathrm{d}r \left[  \left(  1-\frac{b(r)}{r}\right)  ^{-1}\frac{1}{r^{6}%
}\left(  C_5+2\int \mathrm{d}r \ r^6 V^{\prime}(r)\right)  \right]  ^{\frac{1}{2}} +\psi_{0} \,
\end{equation}
where $C_5$ is an integration constant. Unfortunately, there is no analytical general solution for this case.

\section{Conclusions}
\label{sec:concl}

In this paper we have taken under examination the combined contribution of a
scalar field and a Casimir source for the formation of a TW. First, we considered a scalar field
without a potential, and second, we examined the case with a potential. In addition, the  distance of the Casimir plates has been considered either to be
 parametrically fixed or  radially variable. In
every profile, a solution appears but it must be confined in a bounded region
to avoid divergences on the redshift or the shape function. This means that our model does
not satisfy the asymptotic flatness, but it has a finite range of definition.
However, this may not pose a problem, as it is always possible to use the junction method (thin shells) to connect the inner core of the traversable wormhole with a Minkowski metric. We intend to address this issue in future research. In addition, to establish a
consistent system of equations, we had to consider the introduction of a third 
field $T_{\mu\nu}^{\mathrm{TW}}$, which is interpreted as the gravitational response of
the TW to the action of the matter fields $T_{\mu\nu}^{\mathrm{Casimir}}$ and
$T_{\mu\nu}^{\mathrm{Scalar}}$.
Nevertheless, when considering a scalar field with a potential term in the presence of a Casimir source with fixed plates, the introduction of $T_{\mu\nu}^{\mathrm{TW}}$ is unnecessary. It is also worth noting that, in the case without a potential, the predicted size of the throat is enormous, since it is proportional to the ratio  $d^{2}/r_{1}\simeq1.7\times10^{17}m$. At this stage of our research, we are unaware of the reasons behind this result.

\end{document}